# Investigating the completeness and omission roads of OpenStreetMap data in Hubei, China by comparing with Street Map and Street View

Qi Zhou [1,2,*], Hao Lin [1]

| 1 | School of Geography and Information Engineering, China University of Geosciences, Wuhan, P.R. China |
| 2 | Beijing Key Laboratory of Urban Spatial Information Engineering, Beijing, P.R. China |

Email: zhouqi@cug.edu.cn | cug_bird@cug.edu.cn

**Abstract:** OpenStreetMap (OSM) is a free map of the world which can be edited by global volunteers. Existing studies have showed that completeness of OSM road data in some developing countries (e.g. China) is much lower, resulting in concern in utilizing the data in various applications. But very few have focused on investigating what types of road are still poorly mapped. This study aims not only to investigate the completeness of OSM road datasets in China but also to investigate what types of road (called omission roads) have not been mapped, which is achieved by referring to both Street Map and Street View. 16 prefecture-level divisions in the urban areas of Hubei (China) were used as study areas. Results showed that: (1) the completeness for most prefecture-level divisions was at a low-to-medium level; most roads (in the Street Map), however, with traffic conditions had already been mapped well. (2) Most of the omission OSM roads were either private roads, or public roads not having yet been named and with only one single lane, indicating their lack of importance in the urban road network. We argue that although the OSM road datasets in China are incomplete, they may still be used for several applications.

## 1.Introduction

Along with the development of Web 2.0 technology, an ever-increasing amount of geographic information data has been created and updated by volunteers, a practice called "volunteered geographic information"[1]. As one of the most successful examples, OpenStreetMap (OSM) is an online map (http://www.openstreetmap.org/) which can be edited and updated by volunteers all over the world. The OSM data is not only free to use but also has a global coverage. OSM can be an essential data source to enhance digital earth products[2], and it also has the potential to play an important role in digital earth applications[3], a wide range of which include 3D modeling[4], path planning[5], emergency relief[6], and land use and cover mapping[2,7]. These applications are based on OSM data despite concerns relating to the quality of the data arising from most OSM volunteers being non-specialists and amateurs[1,8].

Extensive research has focused on assessing various quality elements of OSM data, including positional accuracy, thematic accuracy, topological consistency, and completeness[8-11]. Among these different quality elements, the matter of completeness (deciding whether a region has been well covered) has gained much attention. Haklay assessed the completeness of OSM road datasets in England by comparing them with a corresponding authorized dataset – Ordnance Survey (OS)[8]. He used the difference between the road lengths in the OSM and OS datasets as a completeness measure, a measure which happens also to have been used for assessing the completeness of OSM road datasets in the United States[12]. Koukoletsos et al. proposed an automated method for matching OSM and OS datasets, calculating the completeness in terms of the lengths of matched roads proportional to the total length of roads in either the OSM or OS dataset[13]. Ludwig et al. used a similar method to compare OSM and Navteq (produced by a commercial company – HERE) road datasets in Germany[14]. Girres and Touya used a ratio of road lengths between OSM and authorized datasets in France as the completeness measure[9], while Brovelli et al. used the length percentage of an OSM road dataset included in a predefined buffer of a reference dataset[15]. Ciepluch et al. compared three typical online maps of Ireland (OpenStreetMap, Google Maps, and Bing Maps) by manually counting the number of errors (e.g. "incorrect streetname", "incorrect road or street designation")[16]. Common to all these studies is a reference dataset (obtained from either a mapping agency or commercial company) for assessing the completeness of an OSM dataset. Such a reference dataset is not always available, however, due its expense or being out of date. Thus, several attempts have been made to assess the quality of OSM data without reference datasets, a practice known as "intrinsic quality assessment." Barro et al. proposed a framework including 25 indicators for OSM quality assessments based solely on the data's history[17]. Neis et al. analyzed the evolution of the OSM road dataset in Germany from 2007 to 2011[18]. Neis et al. selected 12 urban areas from around the world and found a correlation between socio-economic factors (e.g. income) and data provided from the analyzed areas[19]. Gröchenig et al. analyzed the annual changes of OSM features and defined three stages: Start, Growth, and Saturation, which described a certain degree of data completeness in the analyzed regions[20]. Zhou and Tian proposed three indicators, i.e. street block area, perimeter, and density to quantitatively estimate the street block completeness of OSM road datasets; a street block denoted a closed region formed from several road segments[21]. Sehra et al. developed a toolbox based on an open-source geographic information system (GIS) software package, Quantum GIS, to perform intrinsic quality assessment[22].

Furthermore, the above studies showed that OSM data is almost complete in developed countries or regions. For example, Neis et al. found that the OSM road dataset in Germany had a small difference (only 9%) compared with a commercial dataset for car navigation[18]. Barrington-Leigh and Millard-Ball also reported that more than 40% of countries (e.g. United States, Japan, France, Canada, Australia, and Germany) had a fully mapped OSM road dataset, but that the completeness was much lower in some other countries (e.g. China, Russia, and India)[23]. Ming et al. assessed the OSM road dataset in Wuhan (a prefecture-level division of China) and found that the road completeness of this region was only 38%[24]. Zhou et al. also analyzed some urban regions in China and found that the OSM road completeness varied from 28.61% to 60.77%[25]. Zheng and Zheng assessed the OSM road dataset in China by comparing it with the dataset produced by Baidu (a commercial mapping company in China), finding that 71% of the OSM data was less detailed than the Baidu data; more than 94% of the country consisted of incomplete regions[26]. Tian et al. analyzed the completeness of OSM building data in China and concluded that the building data were also far from complete[27].

The aim of this study is to reinvestigate the completeness of OSM road data in China because firstly, in China, most geographical data produced by either mapping agencies or commercial mapping companies have not been made publicly available, and it is desirable, therefore, to obtain open-source data as a supplement. OSM data, being freely available, may be used as an alternative. Second, as one of the most essential sources of data, road data can be used for many applications, including traffic flow prediction, map representation, navigation, and routing[5,28,29]. Nevertheless, most existing studies report that the completeness of OSM road data in China is relatively much lower than that of the data in developed countries, resulting in serious concern in utilizing Chinese OSM road data in applications. More importantly, to our knowledge, very few studies have focused on investigating what types of road are still poorly mapped in the OSM road dataset of China.

This study investigates not only the completeness but also the omission roads (e.g. the unmapped or omitted roads) of OSM data in China.

This study makes two main contributions: First, a hierarchical classification scheme is proposed to analyze omission roads in OSM by referring to both Street Map and Street View. This is especially necessary for countries (e.g. China) whose reference datasets are not freely available. Second, we found that most omission roads in the urban areas of China are the least important roads in an urban road network, being either private roads, or public roads unnamed and consisting of only a single lane. We argue that the existing OSM road datasets in the urban areas of China, although incomplete, may still be used for several applications.

This study is structured as follows: Section 2 designs a series of experiments to investigate both the completeness and omission roads of the OSM road data in China. Section 3 reports the experimental results and analyses, while sections 4 and 5 present the discussion and conclusions, respectively.

## 2. Design of experiments

### 2.1 Study area and data

We studied the urban areas of the different prefecture-level divisions of Hubei province, China (Figure 1), a choice of areas predicated on the following: First, previous studies have verified that both population and social-economic factors potentially affect the completeness of an OSM dataset[19,27]. The Hubei province, located in the central region of China, was ranked 9th and 7th in 2018 among the 34 provinces of China, in terms of population and gross domestic product (GDP), respectively; this province indicates a medium level of the OSM dataset developed in China. Second, there are a number of prefecture-level divisions in Hubei (17), which may minimize the subjectivity of using only a division as the study area. More importantly, in this study, the Baidu Maps were employed as a reference map against which the OSM dataset was compared, and the Baidu Street View, as one of the functions in Baidu Maps, was also employed for analyzing the omission roads of the OSM data. However, the Baidu Street View was only available in most of the urban areas rather than in rural areas, and thus only the urban areas in those prefecture-level divisions of Hubei were analyzed. To be specific, the datasets and maps used in this study are as follows:

- OSM road dataset: The OSM road dataset in Hubei (China) was downloaded from the website: http://download.geofabrik.de/asia/china.html in Jan. 2019, and extracted from the whole country.
- GLOBALAND30 (GLC30) dataset: The GLC30 dataset is a global land cover data in 2010 at 30-meter resolution. This dataset, which can be freely obtained from the website: http://globallandcover.com/GLC30Download/index.aspx, was produced by the National Geomatics Center of China. The class "artificial surface" in the GLC30 dataset was viewed as urban areas.
- Baidu Maps: Baidu Maps is an online mapping application developed by the Baidu company offering not only a street map (Baidu Street Map) but also a street view perspective (Baidu Street View). Both Baidu Street Map and Baidu Street View were used for analyzing the completeness and omission roads of OSM road data. In addition, we found that there were very few roads mapped in the OSM road datasets of China but none in the Baidu Street Map. This illustrates that Baidu Maps, which are much more complete, can be used as a reference.

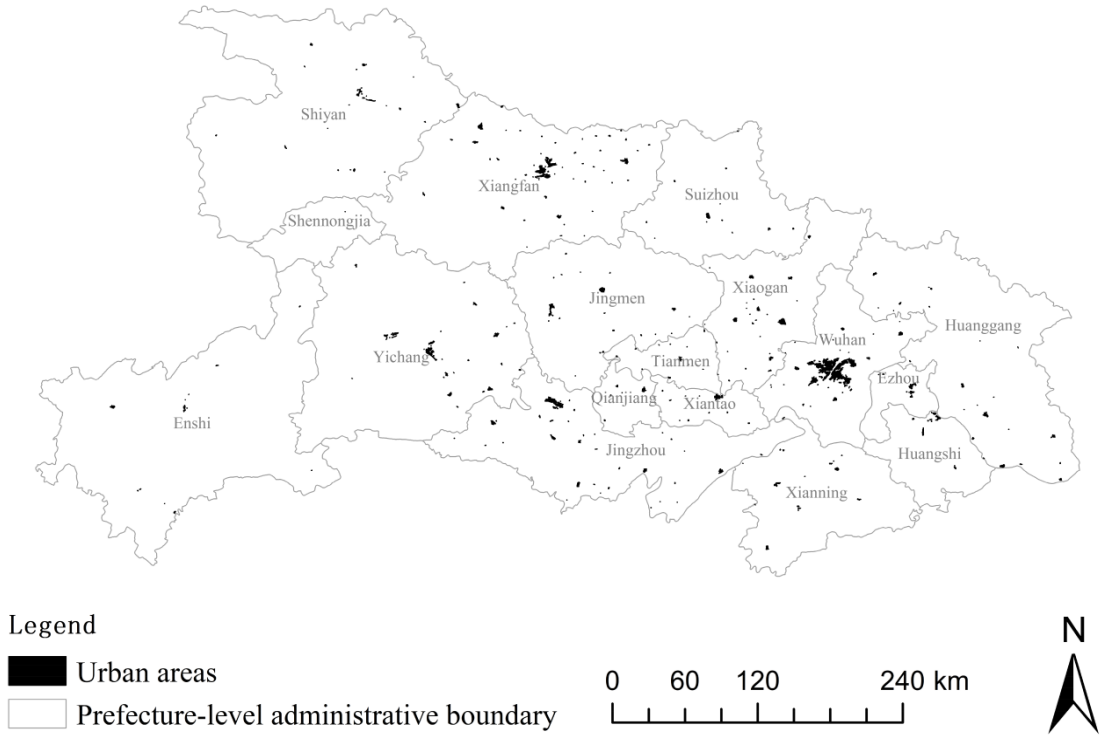

Figure 1: Study area.

## 2.2 Methods

(1) Analysis of OSM road completeness

Much research has focused on the use of a corresponding reference dataset to compare it with an OSM dataset. However, such a reference dataset for our study area was not freely available; therefore, we employed the approach proposed by Zhou and Tian[21] to assess the road completeness in OSM. This approach sought to analyze the completeness of street blocks in an OSM road dataset by comparing them with a reference map. Here, a street block was defined as a closed region (e.g. A' and B' in Figure 2(a)) surrounded by several road segments. More precisely, Figure 2 was used to explain this approach.

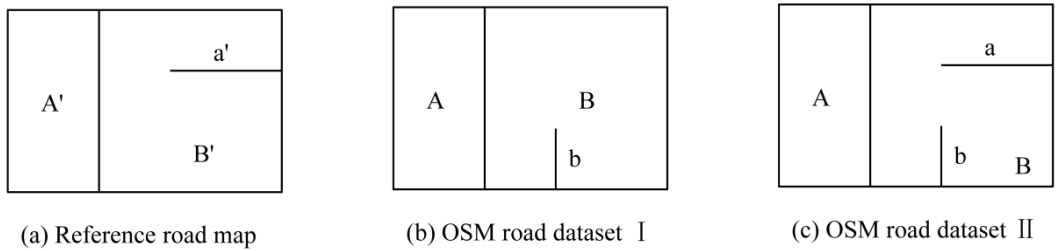

(a) Reference road map     (b) OSM road dataset Ⅰ     (c) OSM road dataset Ⅱ

Figure 2: A series of schematic road networks to explain how to determine the completeness of a street block.

Figure 2(a) shows a reference road map, which includes two street blocks (A' and B') and a road segment a' inside street block B'. Figures 2(b) and 2(c) show two corresponding OSM road datasets with different cases (I and II).

- In the OSM road dataset I (Figure 2(b)) there are two street blocks, A and B, and one

road segment, b, inside street block B. Street block A is viewed as complete because there is no additional road segment inside corresponding street block A' in Figure 2(a). However, street block B is viewed as incomplete despite its additional road segment, b. This is because road segment b is not matched with anything through visually comparing it with (a') in Figure 2(a). The additional road segment a' in Figure 2(a) was viewed as an omission road of the OSM road dataset I.

- In the OSM road dataset II (Figure 2(c)) both street blocks A and B are viewed as complete because road segment a visually corresponds to a' in Figure 2(a); moreover, there is no additional road segment inside corresponding street blocks A' and B' in Figure 2(a).

Additionally, many online mapping applications (e.g. Baidu Maps and Google Maps) provide traffic conditions for major roads in a city (Figure 3). In this study, the completeness of each street block was also determined by considering only roads with traffic conditions; that is, an OSM street block was determined as complete if, inside the corresponding street block in the reference map, there was no additional road segment with traffic conditions. For example, in Figure 2(b), street block B is viewed as complete if road segment a' in Figure 2(a) does not show any traffic conditions. The purpose of this analysis is to investigate whether major roads have already been mapped in an OSM dataset.

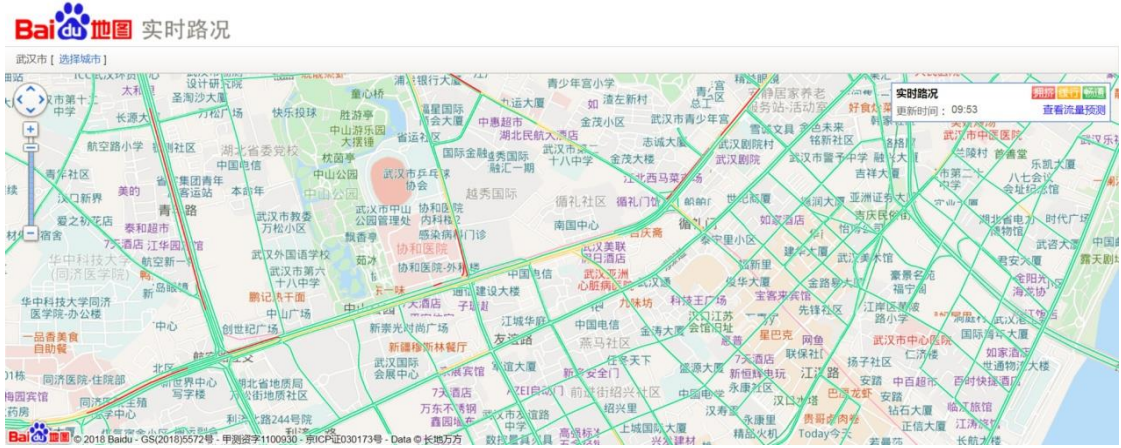

Figure 3: A screenshot of roads with traffic conditions, highlighted with red, yellow, and green, in the Baidu Street Map (©2019 Baidu).

After determining the completeness of each street block in an OSM road dataset, the completeness for a number of street blocks (denoted as the street block completeness) can be calculated. That is, the street block completeness ($C_{street\ blocks}$) measures the percentage of the number of complete street blocks ($N_{complete\ street\ blocks}$) proportion to the number of all street blocks ($N_{all\ street\ blocks}$) in a road network (equation (1), Zhou and Tian[21]).

$$C_{street\ blocks} = \frac{N_{complete\ street\ blocks}}{N_{all\ street\ blocks}} \times 100\% \tag{1}$$

In this study, two cases of completeness values were calculated. For the first case (all roads), all roads in a reference map were considered to determine the street block completeness; and for the second case (roads with traffic conditions), only roads (in a reference map) with traffic conditions were considered.

(2) Analysis of omission roads

This method seeks to design a hierarchical classification scheme to analyze the omission roads in each street block (Figure 4). This is because several attributes (e.g. road name, number of lanes, and road function) of an omission road may be visually determined by referring to the Baidu Street Map and/or Baidu Street View. More precisely, this scheme includes two levels.

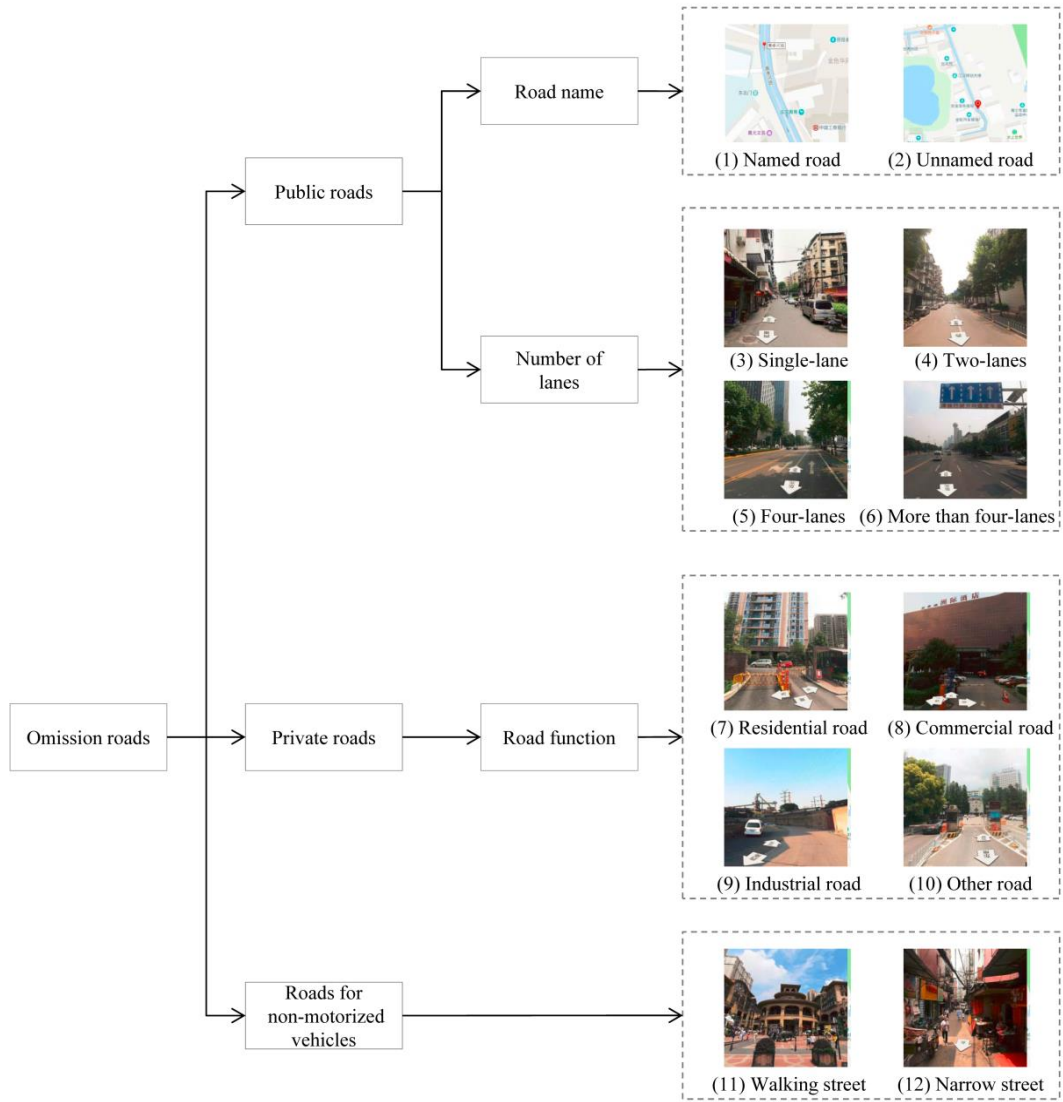

Figure 4. A hierarchical classification of omission roads.

At the first level, an omission road is classified into one of three types: public roads, private roads, and roads for non-motorized vehicles, which may be visually determined by referring to the Baidu Street View. That is:

- "Public road": A road owned and/or maintained by a public authority (e.g. a municipality) and open to public traffic. In an urban road network of China, almost all the four road types: expressway, main road, secondary road, and branch road can be divided into public roads. The main criterion to determine a public road is that it is open to public traffic (see photos (3)–(6) in Figure 4).
- "Private road": A road owned and/or maintained by a private individual, organization, or company, rather than a government. A movable barrier is often used to restrict public traffic access to a place, such as a gated community or factory. The main criterion to determine a private road is that it is not open to public traffic, or there is a barrier to restrict public traffic access (see photos (7)–(10) in Figure 4).
- "Road for non-motorized vehicles": A road only used for non-motorized vehicles (e.g. bicycles and/or pedestrians) rather than motorized vehicles. The main criterion to determine this type of road is that such a road is a street for walking, or it is too narrow for public traffic access (see photos (11)–(12) in Figure 4).

At the second level, each omission road, marked as either a public or private road, is further divided into different sub-classes:
- "Public road": We investigated whether a name existed for the corresponding road in the Baidu Street Map and then such a road was divided into one of two types: "named road" and "unnamed road" (see photos (1)–(2) in Figure 4). Moreover, the number of lanes for this type of road was counted by referring to the Baidu Street View, and such a road was further divided into one of four types: single-lane, two-lanes, four-lanes, and more than four-lanes (see photos (3)–(6) in Figure 4).
- "Private road": Each road was marked as one of the four (land-use) functions: residential, commercial, industrial, and others. The Baidu Street View was used to divide private roads into different sub-types (see photos (7)–(10) in Figure 4). For example, a private road inside a residential community is sub-divided into a residential road.

## 2.3 Experimental steps

(1) Analysis of OSM road completeness
- Step 1: Extract the OSM road datasets in the urban areas of different prefecture-level divisions in Hubei; convert the OSM road dataset of each prefecture-level division into a number of street blocks.
- Step 2: Determine the completeness of each street block through visually comparing it with the Baidu Street Map. When using this reference map, consider two cases (all roads and roads with traffic conditions).
- Step 3: Count the number of complete street blocks and calculate the completeness of street blocks in each prefecture-level division according to equation (1). For each prefecture-level division, calculate completeness values respectively for the two cases in Step 2.

(2) Analysis of omission roads
- Step 4: Randomly pick up 60 incomplete street blocks from the OSM road dataset in each prefecture-level division.
- Step 5: Overlap each of the 60 street blocks with the corresponding map in the Baidu Street Map; manually digitize all the omission roads in each of these street blocks. As it was a time-consuming process to digitize roads from the Baidu Street Map, in each prefecture-level division, only 60 street blocks were chosen as samples.
- Step 6: Visually determine various types (in Figure 4) of each omission road by referring to both the Baidu Street Map and Baidu Street View; calculate the total length of omission roads for each type.

## 3. Results and analyses

### 3.1 Results of OSM road completeness

Figure 5 plots the street block completeness for the urban areas in the 16 prefecture-level divisions of Hubei (China), considering two cases (all roads and roads with traffic conditions). Shennongjia was excluded from the analysis because there were only six urban street blocks in this prefecture-level division.

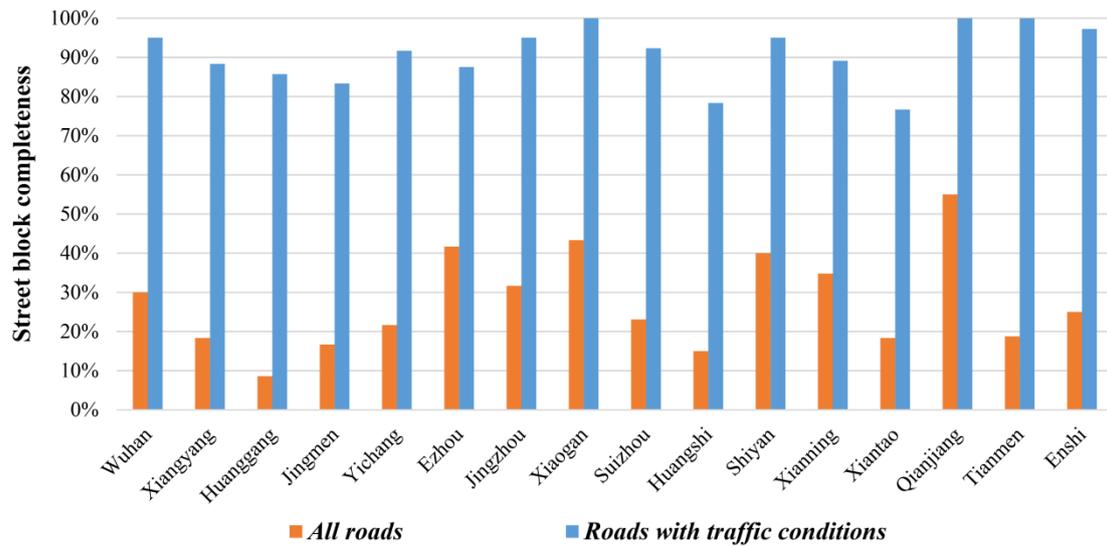

Figure 5: Street block completeness values for the urban areas in the 16 prefecture-level divisions of Hubei, China.

Figure 5 shows that with respect to all roads, for 13 out of the 16 prefecture-level divisions, street block completeness values were lower than 40%, and the maximum value was only 55% (Qianjiang), illustrating a lack of roads in the OSM road dataset of China. This finding is also consistent with that found by other studies[21,24,25]. With respect to roads with traffic conditions, however, for 14 out of the 16 prefecture-level divisions, street block completeness values were higher than 80%, and those of the others were all close to 80%. This indicates that major roads in these prefecture-level divisions have already been mapped well.

### 3.2 Results of OSM omission roads

Figure 6 plots the length percentages of omission roads for the three road types, i.e. public road, private road, and road for non-motorized vehicles.

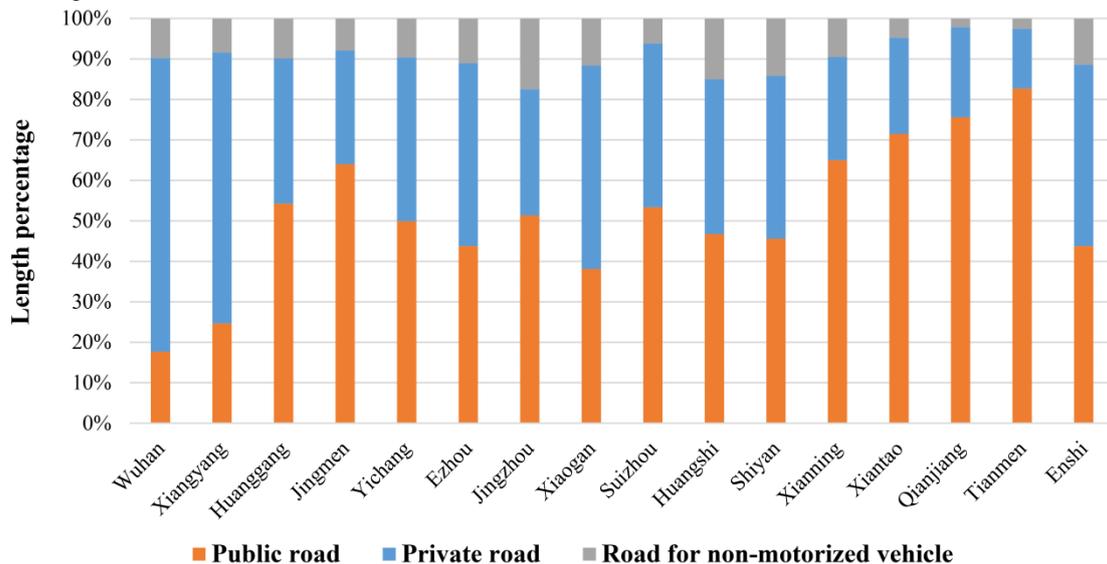

Figure 6: Length percentages of omission roads in the 16 prefecture-level divisions of Hubei (China), in terms of the three road types, i.e. public road, private road, and road for non-motorized vehicles.

Figure 6 shows that in terms of road length, approximately 90% of omission roads were either public roads or private roads, and, in most cases, no more than 10% of them were roads for non-motorized vehicles. This is probably because the Baidu Street Map was mostly mapped for motorized vehicles rather than non-motorized vehicles, and in an urban road network, there are commonly much longer roads for motorized vehicles. Moreover, in 11 out of the 16 prefecture-level divisions, most of the omission roads were public roads rather than private roads, which is evidently not the case for both Wuhan and Xiangyang, in whose prefecture-level divisions there were relatively many more gated communities and/or factories; there was still a lack of both public roads and private roads in the different prefecture-level divisions of Hubei (China).

Figure 7 plots the length percentages of omission roads for different road sub-types in the hierarchical classification scheme (Figure 4) for 16 prefecture-level divisions of Hubei (China). This figure shows:

1) In terms of public roads, more than 60% of omission roads that were mapped in the corresponding Baidu Street Map were unnamed, a percentage which may actually be higher than 90% for both Wuhan and Xiangyang. Such an unnamed road often plays a role in connecting a residential community or commercial facility to a main road (Figure 8). More importantly, more than 90% of omission roads consisted of only a single lane (or single-lane roads), which indicates their lack of importance in an urban road network.

2) With respect to private roads, for 15 out of the 16 prefecture-level divisions, most of the omission roads were residential roads. This is because there were commonly more residential lands in an urban area, and in China there existed a large number of gated communities whose residential roads were private or not open to the public. As a result, probably very few OSM volunteers had paid attention to mapping in these gated communities. This indicates that the private roads were often viewed by OSM volunteers as less important than public roads in an urban road network.

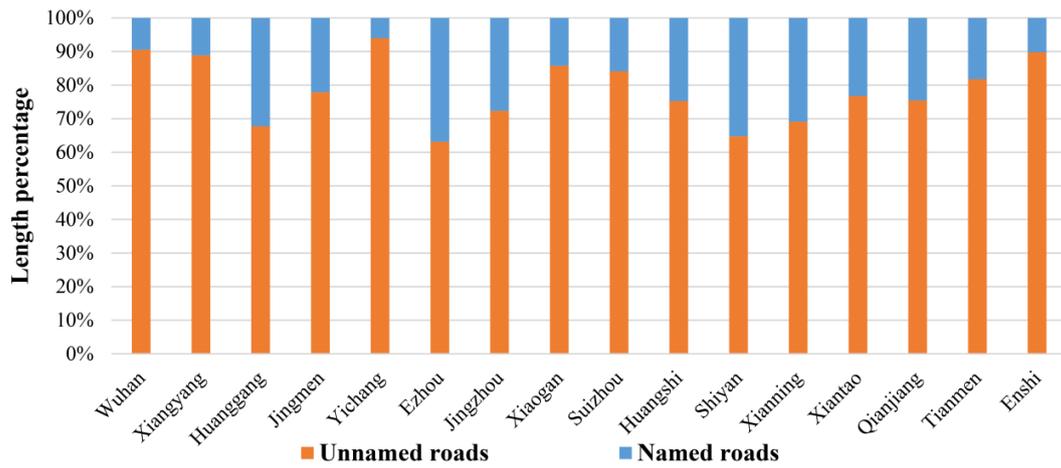

(a) Road name

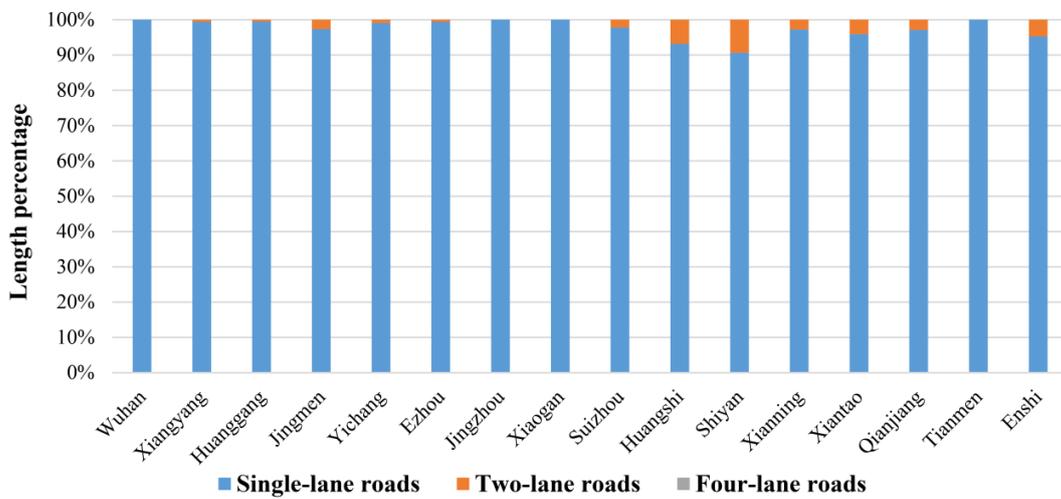

(b) Number of lanes

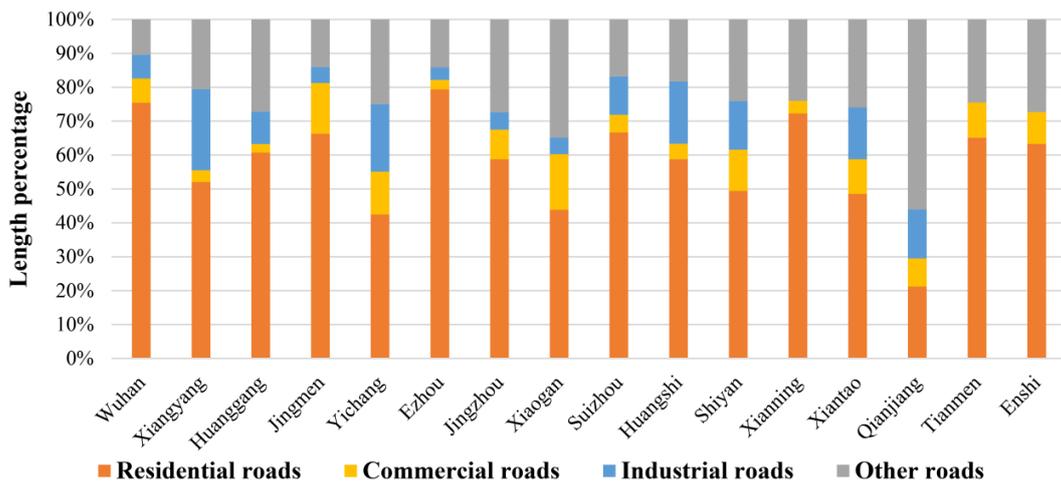

(c) Road function

Figure 7. Length percentages of omission roads for the 16 prefecture-level divisions of Hubei (China), in terms of (a) named or unnamed public roads; (b) public roads with different numbers of lanes; and (c) private roads with different functions.

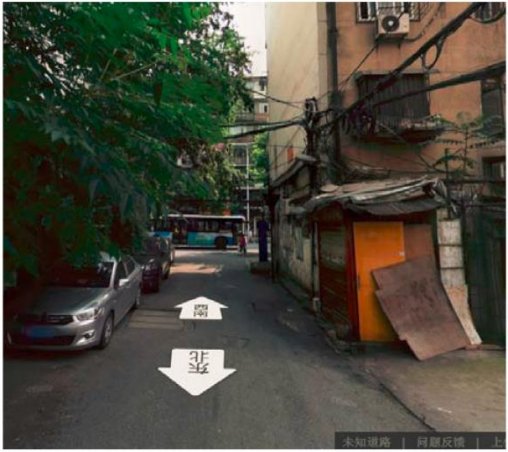 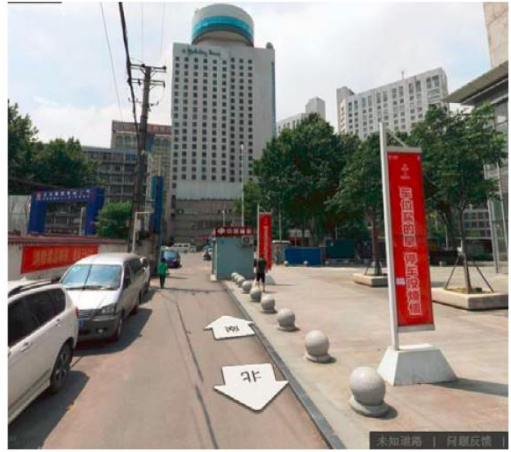

(a) The road connects a residential community to a main road

(b) The road connects a commercial facility to a main road

Figure 8. Omission roads in OSM marked as public roads but not yet named.

## 4. Discussion

From this study, we found that the OSM data datasets in China were indeed not complete, as the completeness values were mostly less than 40% for the 16 prefecture-level divisions of Hubei (China). However, most omission roads (those not existing in the OSM datasets but in the corresponding Baidu Maps) were either private roads, or public roads unnamed and consisting of only one single lane. This means the omission roads in OSM were mostly the least important roads in an urban road network. A similar conclusion is represented by Figure 9, which plots the length of OSM roads in Hubei as a function of the development of OSM road datasets from 2013 to 2019. Two specific scenarios are considered in Figure 9. In scenario I: the total length of the OSM roads completely inside the urban areas of Hubei was calculated. In scenario II, the total length of all the OSM roads inside the administrative division of Hubei was calculated. As an example, eight typical road types, whose road lengths ranked on top, were reported.

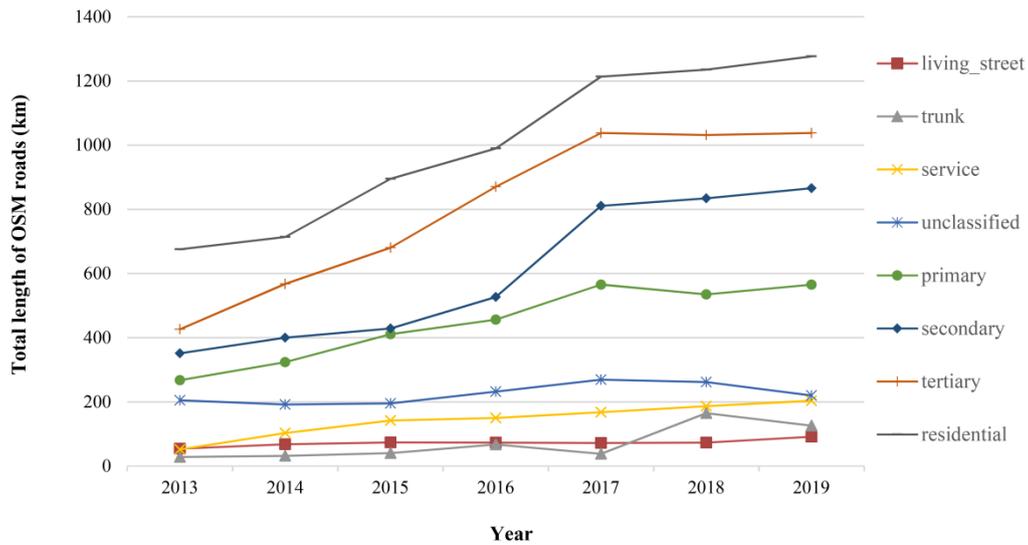

(a) Scenario I

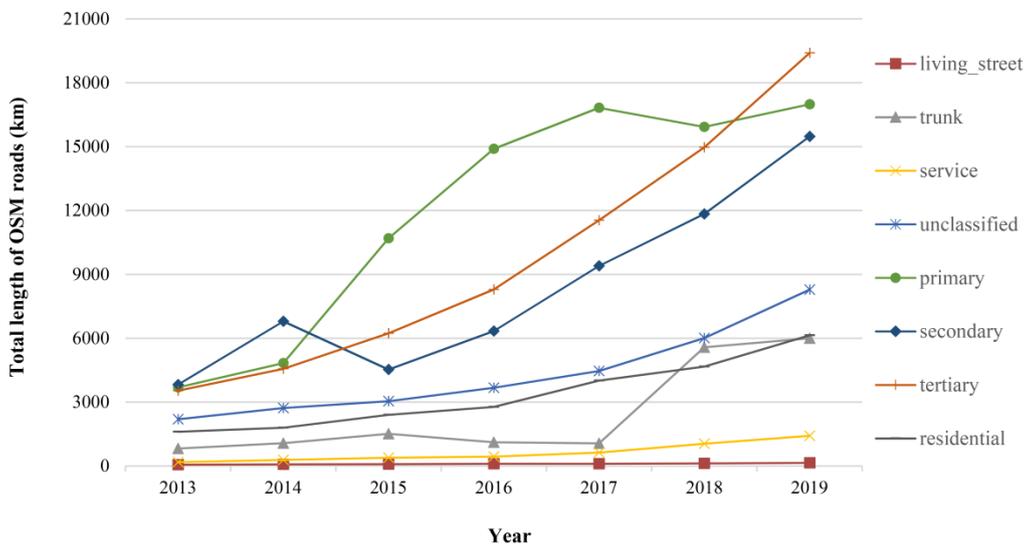

(b) Scenario II

Figure 9. Lengths of OSM roads in Hubei (China) from 2013 to 2019, for eight typical road types. For each year, the corresponding OSM road dataset was acquired in January.

Figure 9 shows that for the first scenario, the lengths of at least four road types ("Primary", "Secondary", "Tertiary", and "Residential") increased rapidly from 2013 to 2017, increasing relatively slowly afterwards. This indicates that relatively important OSM roads (e.g. "Primary" and "Secondary") tend to be complete. For the second scenario, however, a different trend was found, at least for the "Secondary" and "Tertiary" road types. That is, their lengths rapidly increased year-on-year, indicating that these types of OSM road data may not be complete in rural areas, which is consistent with other studies: that is, the OSM data in urban areas were relatively more complete than those in rural areas[9,18]. Nevertheless, it is still necessary to investigate the omission OSM roads in rural areas in future work.

Moreover, some limitations have been found when using the OSM history datasets for analysis. For example, Figure 9(a) shows that the lengths of OSM roads for the type "Trunk" increased sharply from 2017 to 2018; this is because a large number of OSM roads marked as "Primary" had been modified as "Trunk", which resulted in a slight decrease of OSM roads for the "Primary" type (Figure 9(a)). More importantly, this indicates that the road types defined in

OSM were quite different from those used in China. Therefore, a hierarchical classification scheme was designed to analyze omission roads in OSM.

Furthermore, although the OSM road datasets in China are not complete even in the urban areas, they may still be used for a number of applications because most of the relatively important roads (e.g. roads with traffic conditions or public roads with two lanes) have already been mapped well. These OSM road datasets may be used for analyzing traffic flow because in an urban road network, the most important roads ranked in the top 20% can accommodate more than 80% of traffic flow[28]; also, it was found in this study that around 80% of roads (in the reference map) with traffic conditions have already been mapped in the OSM road datasets of Hubei (China). The datasets may also be used for producing a medium- and/or small-scale road network, which is essential for map representation (e.g. in Baidu Maps and Google Maps). In order to produce such a representation, there is a need to eliminate relatively unimportant roads or only retain relatively important roads[29,30]. The OSM road datasets in China may also be useful in understanding the backbone structure of an urban road network, which pays more attention to relatively important streets[31,32]. Nevertheless, as there is still a lack of public roads in the OSM road datasets of China, these datasets may not be used for analyzing an urban road network of a small size or region. Moreover, there is still a need to employ authorized datasets for assessing the usability of the OSM road data in China for various applications.

## 5. Conclusion

This study carried out a series of experiments to investigate both the completeness and omission roads of OSM road datasets in China. The completeness for a number of OSM street blocks in the urban areas was determined by comparing the blocks with the Baidu Street Map, for which two cases (all roads and roads with traffic conditions) were considered. A classification scheme was designed to investigate what types of roads were omitted in the randomly sampled street blocks, determined through comparing them with both the Baidu Street Map and Baidu Street View. Sixteen prefecture-level divisions of Hubei province (China) were involved for the analyses. Results showed the following:

1) In terms of all roads, the completeness value for most of the prefecture-level divisions was less than 40%, indicating that even in urban areas, the OSM road datasets of China were not complete. In terms of roads with traffic conditions, however, 80% or more street blocks were complete in most cases, indicating most major roads had already been mapped well.

2) Furthermore, most of the omission OSM roads in the urban areas were either private roads (mostly residential roads), or public roads unnamed and consisting of only a single lane, indicating that most omission OSM roads were the least important roads in an urban road network.

This means that the OSM road datasets in China, although incomplete, may still be used for several applications (e.g. traffic flow analysis, road map representation, and backbone structure detection) as discussed in Section 4. Further research may include: (1) verification of the availability of OSM road datasets in China for specific applications; (2) investigation of the omission OSM roads in rural areas; and (3) validation of the above results with respect to other regions in China.

**Funding:** This project was supported by the National Natural Science Foundation of China (No. 41771428), Fundamental Research Funds for the Central Universities, China University of Geosciences (Wuhan), and Beijing Key Laboratory of Urban Spatial Information Engineering (No. 2019205).

**Acknowledgments:** The authors want to thank the OpenStreetMap © contributors for making available as open their databases.